# Pure temporal dispersion for aberration free ultrafast time-stretch applications

**Liao Chen, Xin Dong, Ningning Yang, Lei Zhang, Xi Zhou, Zihui Lei, Chi Zhang, [*] and Xinliang Zhang [*]**

*Wuhan National Laboratory for Optoelectronics, Huazhong University of Science and Technology, Wuhan, 430074, China*

**Photonic time-stretch overcomes the speed limitation of conventional digitizers, and enables the observation of non-repetitive and statistically rare phenomena that occur on short timescales. In most of the time-stretch applications, large temporal dispersion is highly desired to satisfy the far-field diffraction regime. However, most conventional spatial disperser or chirped fiber Bragg grating are constrained by its spatial volume, which can be overcome by the ultra-low-loss dispersive fiber, as an ideal medium for large temporal dispersion ($β_2$), while it suffers from the third-order dispersion ($β_3$) and aberrations. In this paper, an optical phase conjugation based third-order dispersion compensation scheme is introduced, with accumulated $β_2$ and eliminated $β_3$, and achieved ±3400-ps² pure temporal dispersion of over 30-nm bandwidth. Leveraging this pure**

**temporal dispersion, up to 2% temporal aberrations have been eliminated; furthermore, a Fourier domain spectroscopy has achieved a record 15000 optical effective resolvable points, with non-degraded 2-pm resolution over 30-nm range.**

Temporal resolution of the optical measurement technologies has developed by leaps and bounds, no matter the spatial array detector with 500-ns resolution and over a thousand pixels [1], or the temporal single-pixel detector with ultrafast 10-ps resolution [2]. Especially the latter single-pixel detector, united with the dispersive time-stretch which stretches the spectral information along the time axis, catches the researchers' attentions, due to its superior overall data volume [3,4]. This time-stretch based spectroscopy has been applied in a variety of applications and unveiling fascinating ultrafast phenomena, including femtosecond digitizer [5,6], discovery of optical rogue wave [7], ultrafast microscopy [8,9], soliton explosions and molecules [10,11], stimulated Raman spectroscopy [12,13], and mode-locking process [14,15]. All these applications promote the grade of the ultrafast real-time measurements with exceeding bandwidth, accuracy, and record length, and enable the observation of ultrafast non-repetitive and statistically rare signals [3]. Besides the dispersive time-stretch, the other essential part of the temporal optics is the time-lens, which further enhances the manipulation between the frequency and time domains [16,17,18]. Consequently, a number of temporal imaging systems have been developed, including temporal magnification and compression [19,20], temporal Fourier transformation [21,22], temporal filtering [23], temporal convolution [24,25], and temporal cloaking [26,27]. Ideally, aforementioned ultrafast time-stretch applications can perform the designed characteristics; however, owing to third-order dispersion of the dispersive fiber, it will introduce nonlinearity to frequency-to-time mapping, or higher-order phase terms in the time-lens [28]. Therefore, this third-order dispersion will also introduce aberrations to the temporal systems. Take the temporal magnification for example, its time-bandwidth product (TBWP), which is the ratio between the recording length and the temporal resolution, has not exceed 450, due to the existence of third-order dispersion [21]. To overcome its constraints to the resolution and bandwidth, direct studies of the optical temporal dispersion are highly desired.

Temporal dispersion, also known as group velocity dispersion (GVD), optically separates different wavelength along the time axis, and enables the instantaneous frequency-to-time mapping and single-shot measurement [3]. Its implementation starts from the spatial disperser, e.g. prism or diffraction grating pairs, by introducing gradually changed delay along the dispersed spatial direction, though the third-order dispersion exists, it has been widely applied in the femtosecond pulse compression [29]. However, the spatial dispersers have small dispersion or bandwidth, due to limited spatial volume, and it is not enough for the far-field (Fraunhofer) diffraction regime to perform the frequency-to-time mapping [18]. Leveraging the multiple reflections between angle-misaligned mirrors to enlarge the spatial volume, known as the free-space angular-chirp-enhanced delay (FACED), the temporal dispersion can be as large as 1 ns/nm, and successfully applied to the ultrafast time-stretch microscopy [30]. Besides the spatial disperser, the optical fiber, with reduced degree of freedom and ultra-low-loss (e.g. 0.14 dB/km), becomes an ideal component to achieve large temporal dispersion [31]. With the rapid development of the optical fiber communication, different types of fiber have been developed with versatile functions, such as the G.652 signal-mode fiber (SMF) with dispersion of –22.7 ps$^2$/km (or GVD of 17.8 ps/nm/km), the dispersion compensation fiber (DCF) with dispersion of 146.6 ps$^2$/km (or GVD of –115 ps/nm/km), and the dispersion-shifted fiber (DSF) with almost zero dispersion (all at 1550-nm wavelength range) [32]. Among them, due to the larger dispersion-to-loss ratio, DCF is the most commonly used temporal dispersion medium in the time-stretch applications [3,8]. However, the existence of the third-order dispersion coefficient (e.g. 0.113 ps$^3$/km in the SMF, –0.84 ps$^3$/km in the DCF) is easily overlooked, and it results in the nonlinearity of the frequency-to-time mapping process, as shown in Figs. 1(a) and 1(b). Although some specially designed fibers or photonic crystal fibers have been simulated to achieve large dispersion and minimum third-order dispersion, they are not commercially available due to the complex design and fabrication process [33,34].

To eliminate the third-order dispersion of optical fiber, an alternative solution is the chirped fiber Bragg grating (CFBG), with Bragg reflectors distributed along the fiber direction, and it can achieve pure temporal dispersion [35]. It is noted that the round-trip delay time of the CFBG is actually the maximum dispersive delay, which limits the product of dispersion and bandwidth. Due to the limited phase mask size, the typical length of conventional CFBG is ~1-3 cm, and over 2-m length CFBG has been reported based on high-precision translation stage, with –2.7-ns/nm dispersion and 10-nm bandwidth [36]. However, due to the imperfections in the fabrication process, the CFBG gives rise to group-delay ripple, which creates unwanted satellite pulses and intensity fluctuations, and it is harmful to the time-stretch applications [37]. In view of these constrains, another potential scheme to implement the pure temporal dispersion is to compensate the third-order dispersion between two components. Take the SMF and DCF for example, their third-order dispersion coefficients are opposite, and can be totally eliminated by precisely matching fiber length. However, it is noted that the overall temporal dispersion of the SMF and DCF is also compensated by and large, and results in poor dispersion-to-loss ratio [32].

Recently, B. Li et. al. introduces the large effective-area fiber (LEAF) to compensate the third-order dispersion of DCF, and performs the pure temporal dispersion in the temporal magnification system [38]. Since the LEAF has similar third-order dispersion coefficient (0.1268 ps$^3$/km) as SMF, and negligible GVD (–3.6 ps$^2$/km), the overall GVD of the DCF and LEAF is mainly determined by that of the DCF. Although this scheme helps the temporal magnification system to achieve over 2000 record TBWP, the third-order dispersion of output dispersion is still un-compensated, owning to the large dispersion (−1.689 ns/nm) requires over 100-km LEAF to compensate its third-order dispersion. In other words, this scheme is only suitable to compensate the third-order dispersion of short DCF (e.g. < 2 km).

In contrast to the direct-cascaded dispersive compensation, which is the sum of all the dispersion coefficients, the optical phase conjugation (OPC) suggests a new way of thinking about the third-order dispersion compensation [39]. It is noticed that, the temporal complex conjugation would result in the frequency complex conjugation with the variable changed in sign, namely the equivalent frequency-domain transfer function $G(\omega)$ becomes $G^*(–\omega)$, where the asterisk represents the complex conjugation [40]. In the dispersive time-stretch process, the frequency-domain transfer function can be expressed as: $G(\omega) = \exp(– i\beta_2 L\omega^2/2 – i\beta_3 L\omega^3/6 – i\beta_4 L\omega^4/12 –...)$, and after the OPC process, the even-order dispersion coefficients (e.g. $\beta_2$, $\beta_4$) reverse sign, while the odd-order dispersion coefficients (e.g. $\beta_3$) keep identical. As we know, the OPC based dispersion (second-order) compensation has been applied in the optical fiber communication system, with the identical SMFs placed before and after the OPC [41]. While in the time-stretch applications, large pure temporal dispersion is required, thus it is required to be configured in a different way, with the combination of SMF and DCF instead. Leveraging the OPC process, which only reverses the sign of $β_2$, thus the $β_2$ is accumulated while the $β_3$ is eliminated, as shown in Fig. 1(d). If they are direct-cascaded, their opposite $β_2$ and $β_3$ coefficients will be eliminated. This OPC scheme enables the implementation of pure temporal dispersion with large dispersive temporal delay.

In this paper, this pure temporal dispersion was implemented through the four-wave mixing (FWM) based OPC, with 80-km SMF and 10.8-km DCF, and it achieved the overall temporal dispersion of ±3400 $ps^2$ (~ 2.67 ns/nm) over 30-nm bandwidth (from 1527 nm to 1557 nm, or 1565 nm to 1595 nm). Meanwhile, the third-order dispersion was compensated from –19.524 $ps^3$ (equivalent DCF) to –0.012 $ps^3$, which is three orders of magnitude improvement, thus it can be treated as the pure temporal dispersion. Moreover, this scheme can easily switch between the normal dispersion and anomalous dispersion, by swapping the positions of the DCF and SMF. To investigate the third-order dispersion introduced temporal aberrations, this pure temporal dispersion, as a comparison to the conventional DCF, has been applied in a variety of ultrafast time-stretch systems. In the linear domain dispersive Fourier transform (DFT) and time-stretch microscopy, the third-order dispersion introduced 2% nonlinearity was almost compensated, which resulted in finer spectral accuracy, and aberration free imaging [8,12]. While in the Fourier domain, e.g. the swept-source optical coherence tomography (SS-OCT) and the parametric spectral temporal analyzer (PASTA) [22,42], the third-order dispersion introduced severe spectral broadening was also compressed, thus much larger roll-off depth or spectral observation bandwidth has been demonstrated with finer resolution, and a record effective resolvable points (ERP) of 15000 has been achieved in the PASTA system. We believe that this pure temporal dispersion will promote more time-stretch applications, where high frame rate and high accuracy is required.

**Results**

The nonlinearity of the frequency-to-time mapping ratio (Fig. 1(b)) of the dispersive time-stretch is primarily due to the higher-order dispersion, which can be directly characterized from the corresponding pulse delay. Here, the equally spaced (40 GHz over 25-nm bandwidth) frequency components were generated from a wideband pulse source passing through a programmed wavesharper, and the detailed measurement setup is introduced in the supplementary (Section I). Firstly, 23.1-km DCF (OFS, DCM(A)-C-G652) was tested with time-stretched output shown in Fig. 2(a), and its peak positions were reconstructed as the temporal delay of each frequency components (red curve in Fig. 2(d)), whose average slope manifests the second-order dispersion. By removing the linear temporal delay of this second-order dispersion, the residual temporal delay manifests the third-order dispersion, shown as the red parabolic curve in Fig. 2(e). It is noticed that, 1-ns temporal delay of the third-order dispersion introduces 2% nonlinearity for 50-ns overall temporal window. To compensate the third-order dispersion, this 23.1-km DCF was replaced by the OPC scheme with 80-km SMF in the front and 10.8-km DCF afterwards. The results are shown as the blue curves in Figs. 2(b), (d), and (e), especially in Fig. 2(e), the residual temporal delay was compressed to almost zero. In other words, the third-order dispersion is compensated. As a comparison, this OPC scheme was performed in the opposite way by simply switching these two fibers, and output was shown as the black curve in Fig. 2(c). Its temporal delay curve in Fig. 2(d) was fully reversed with the opposite slope and dispersion value, while the residual temporal delay in Fig. 2(e) was still zero.

Directly apply this pure temporal dispersion in the time-stretch microscopy application, where the original aberrations may come from the third-order dispersion, as illustrated in Fig. 1(c) [8]. According to the aforementioned discussion, the nonlinearity introduced by the third-order dispersion of DCF is around 2%, and this misalignment can be viewed as aberration in the time-stretch microscopy, though it is easily overlooked from the real images, which are introduced in the supplementary (Section II). To calibrate this aberration, digitally re-sampling the output time axis is the most straightforward and economical way, while the OPC scheme introduced in this paper provides an alternative optical solution.

Although 2% nonlinearity is negligible in the stretched time domain, it is not the case for its frequency domain counterpart. Here, a Michelson interferometer was introduced to analyze its temporal interference pattern, similar to the SS-OCT system [43]. When the path length difference was gradually scanning, the interference frequency, which is the Fourier transform of the interference pattern, was also scanning accordingly. The intensity response of this frequency scanning process is also called the roll-off curve, e.g. Fig. 3, whose slope reflects the coherence of the swept-source. However, if there are third-order dispersion exists during the time-stretch process, the roll-off frequency will be broadened as the frequency increases, as shown in Fig. 3(a). This broadening is common in the SS-OCT system, and would greatly degrade the imaging performance [42,44]. Therefore, a digital calibration process is usually introduced to compensate the nonlinearity of the swept-source, and the calibrated roll-off curve is shown in Fig. 3(b), with 4.3-mm/dB slope. Details of the calibration process are introduced in the supplementary (Section III). On the other hand, if the third-order dispersion was compensated by the OPC scheme, instead of the digital calibration, almost non-degraded roll-off curve can be obtained as shown in Fig. 3(c), with 10.1-mm/dB slope. In other words, neglect the spatial converging limit, and much deeper tomography depth can be achieved. Moreover, compare the signal-to-noise ratio (SNR) of the roll-off curves in Figs. 3(b) and (c), the OPC scheme achieved 10-dB SNR improvement over the digital calibration, which is probably due to the information loss during the re-sampling of the calibration process. Therefore, this OPC scheme may also result in better imaging sensitivity for the SS-OCT system. Some SS-OCT images are analyzed in the supplementary (Section IV).

To further exaggerate the third-order dispersion introduced aberrations, a temporal focusing spectroscopy, called PASTA, was introduced to perform the ultrafast optical Fourier transform at the temporal focal plane [22,45]. Compared with the aforementioned interferometer, which obtains the interference pattern and performs the digital Fourier transformation, the PASTA system directly obtains the Fourier domain pulses, thus no linear domain pattern available for the digital calibration (detailed setup is introduced in the supplementary Section V) [45]. In pervious works [46,47], the resolution (resolvable spectral width) of the PASTA system is found to be degraded, as it is away from the central wavelength, owning to the third-order dispersion in the output dispersion (quantitative analysis see supplementary Section VI). Therefore, its wavelength observation bandwidth is usually constrained within 5 nm to 10 nm, as shown in Fig. 4(b). Here, the minimum 2-pm resolution (correspond to 2.96-ps pulsewidth) was obtained at central 1540 nm by an asynchronous optical sampling (ASOPS) system (details see in the supplementary Section VII [9]), and it degraded to 15-pm as the wavelength shifted 2 nm to 1538 nm or 1542 nm, matched well with the simulation result (black dotted line) in Fig. 4(c). Replace the DCF with the pure temporal dispersion as the output dispersion, and the wavelength induced spectral broadening was totally removed and 2-pm resolution was achieved for entire observation bandwidth (1527 nm - 1557 nm), as shown in Fig. 4(a). Leveraging this pure temporal dispersion, the temporal focusing based PASTA system achieved the record ERP of 15000 (2000 in the case of real-time acquisition with 33-GHz bandwidth).

Compared with the direct time-stretch based amplified DFT technology [12], which performed as the ultrafast absorption spectroscopy, the temporal focusing based PASTA is advantageous in terms of the observation condition, from the arbitrary noise spectrum, continuous wave (CW) source, to high repetition rate (e.g. > 1 GHz) pulse source [48]. However, previous efforts are mainly limited by its observation bandwidth, which is intrinsic due to the third-order dispersion induced spectral broadening effect. Leveraging this pure temporal dispersion, the spectral broadening effect of the PASTA system was almost removed, and enlarged the observation bandwidth to 30 nm (limited by amplification bandwidth). Deliberate selected signal sources, including the programmed amplified spontaneous emission (ASE) noise, two CW components spaced by 0.05 nm, and 10-GHz pulse source, were successfully resolved by the PASTA with 20-MHz frame rate, which is 4,000,000 times faster than that of the reference optical spectrum analyzer (OSA), as shown in Fig. 5. This ultrafast frame rate for arbitrary signal waveform, in combination with the improved larger bandwidth and finer resolution, provides superior capability for the ultrafast spectroscopy, and it is promising for the spectral dynamic observation of the optical frequency comb from the micro cavity, as well as the ultrafast soliton dynamics [49,50].

**Conclusion**

The concept of time-stretch provides a powerful scientific tool for unveiling ultrafast dynamics, due to the instantaneous frequency-to-time mapping process [3]. Besides its superior frame rate, fine resolution is also achieved in the Fraunhofer regime, with large temporal dispersion [18]. However, due to the existence of the third-order dispersion, the frequency-to-time mapping curve becomes nonlinear, which contributes for the aberrations of the time-stretch applications. Minimum misalignment is firstly observed in the stretched time domain, and it would slightly affect the frequency accuracy rather than its resolution. As it goes to the Fourier domain, the aberrations will be greatly enlarged, with much broadened spectral width and degraded roll-off curve. Moreover, in the time-lens imaging system, such as temporal magnification and temporal focusing [19,22], this nonlinear aberration would constrain its working bandwidth, due to the spectral broadening effect. Therefore, compensate the third-order dispersion would be of great importance for the ultrafast time-stretch applications.

Matching different types of fibers and lengths is the most straightforward way to compensate dispersive parameters, such as SMF and DCF for $\beta_2 = 0$, DCF and LEAF for $\beta_3 = 0$ [32,38]. To achieve $\beta_2 = 0$, due to the large normal dispersion of DCF, short DCF is capable of compensating long SMF [32]. While in terms of $\beta_3 = 0$, the DCF and LEAF scheme is not 'fiber efficient', and much longer LEAF is required to compensate a short DCF, due to the small $\beta_3$ coefficient of LEAF [38]. In other words, without deliberate engineered fiber, this fiber-matching scheme is only suitable for short fiber situation; otherwise large insertion loss will be introduced. As the dispersive fiber length gets longer, in most of the time-stretch applications with Fraunhofer approximation, the OPC scheme will become favorable. It leverages the most conventional DCF and SMF, accumulates their second-order dispersion and subtracts their third-order dispersion, and provides a versatile and efficient dispersion compensation solution.

In this paper, as large as 3400-$ps^2$ pure temporal dispersion has been demonstrated with approaching zero third-order dispersion (~0.012 $ps^3$) over 30-nm bandwidth (from 1527 nm to 1557 nm, or 1565 nm to 1595 nm), based on the OPC scheme. Aforementioned aberrations were thus successfully compensated based on this pure temporal dispersion, from the tiny aberrations in the linear time domain, to the large aberrations in the frequency domain after the interferometer or the optical Fourier transform. Especially in the frequency domain, the compensated time-stretch swept-source has achieved almost non-degraded (1 cm/dB) roll-off curve, which benefits the SS-OCT imaging depth range and sensitivity. For the time-lens based temporal focusing spectroscopy, compensated output dispersion enables 2-pm spectral resolution over 30-nm observation bandwidth with 20-MHz frame rate, which provides a powerful ultrafast spectroscopy for arbitrary signal waveform. Therefore, large pure temporal dispersion is essential for aberration free time-stretch applications.

## Methods

**Time-stretch.** Time-stretch technology employs GVD to transform the spectrum of a broadband optical pulse into a time stretched temporal waveform, and it has been widely applied to wideband analog-to-digital conversion, high-throughput real-time spectroscopy and microscopy, e.g. serial time-encoded amplified microscopy (STEAM) [5,8,12]. The information was first encoded on the spectrum of the broadband pulse, and this spectrum is stretched by large GVD into a sufficiently slowed down temporal waveform, so it can be digitized and processed in real-time. Only consider up to $\beta_2$ and $\beta_3$, its frequency-to-time mapping relation can be simplified as:

$$t = \beta_2 L \Delta\omega + \frac{1}{2}\beta_3 L \Delta\omega^2 \tag{1},$$

where $L$ is the length of the GVD medium. It is obvious that, the existence of the third-order dispersion resulted in the nonlinear mapping relation, and it will definitely result in aberrations for these time-stretch related applications. The OPC scheme introduced in this paper is capable of compensating this nonlinearity, and achieving pure temporal dispersion.

**Optical phase conjugation.** Phase conjugation means the imaginary part of the complex amplitude becomes opposite, and it results in the reversal of the phase property. It is capable of exactly reversing the propagation direction for the spatial wave field, or achieving the time reversal of the temporal field along the optical fiber. The most common way of achieving the phase conjugation in the optical domain is to use the parametric FWM process. For this FWM process, we can describe four electric fields, $E_j$ ($j$ = 1, 2, 3, 4) are the electric field amplitudes. $E_1$ and $E_2$ are known as the two continuous pump waves, with $E_3$ being the signal wave, and $E_4$ being the generated conjugate wave, which can be approximated by:

$$E_4(t) = i\frac{\omega L}{2nc}\chi^{(3)}E_1(t)E_2(t)E_3^*(t) \propto E_3^*(t) \tag{2},$$

where $n$ is the refractive index, $L$ is the interaction length, and $c$ is the speed of light. Considering the $E_1$ and $E_2$ have constant amplitude and frequency profile, the generated idler $E_4$ is the complex conjugate of the signal $E_3$, and this results in the reversal of phase property of the effect. Besides, the conversion bandwidth and efficiency of the OPC process are also critical based on the phase matching condition:

$$-4\gamma P_0 < \Delta k = \beta_4 + \beta_3 - \beta_2 - \beta_1 < 0 \tag{3},$$

where the parametric gain exists. To achieving better observation bandwidth and sensitivity, suitable pump wavelength and power, dispersion engineered nonlinear waveguide are highly desired in this work.

**Roll-off measurement.** The time-stretch swept-source was applied to a Michelson interferometer to implement the roll-off measurement, which is widely applied in the SS-OCT system [46-49]. By scanning the axial position of the reference mirror, the beating frequency of the interference output also proportional changes simultaneously. Therefore, its mapping ratio reflects the chirped ratio of the swept-source, which is 2.67 ns/nm in our measurement. Meanwhile, the peak power and bandwidth degradation of the beating frequency reflect the spectral linewidth and linearity of the swept-source respectively. In view of the broadening of the roll-off frequency, a digital calibration based on *k*-space resampling was applied here, and details see supplementary section III. Consequently, the roll-off slope was improved from 1-2 mm/dB to 4.3 mm/dB, as shown in Figs. 3(a) and 3(b), while the sensitivity was degraded due to the re-sampling process of the digital calibration. As a comparison, the pure temporal dispersion stretched swept-source provided quasi-linear frequency-to-time mapping, thus its roll-off curve was almost flat (10.1 mm/dB slope) without digital calibration.

**PASTA configuration.** The concept and implementation of the PASTA system was aligned with Ref. [22,45], expect the output dispersion has been updated with this OPC introduced pure temporal dispersion, see supplementary Section V. The converging time-lens was implemented with the parametric FWM process in the highly-nonlinear fiber (HNLF), and its swept pump had the dispersion of 2.96 ns/nm. The conversion efficiency of the parametric process was –10 dB, with averaged pump power of 20 dBm. Filtered idler of the parametric process passed through the output dispersion, where the spectral resolution was quite sensitive to its higher-order dispersion. Detailed analysis of the higher-order dispersion in the pump dispersion and output dispersion is introduced in the supplementary (Section VI). Afterwards, the temporally resolved spectral pulses were acquired by the acquisition system with 33-GHz bandwidth. Alternatively, the ultrafast temporal pulses were also analyzed by the ASOPS system to pursue the finest spectral resolution (up to 2 pm, as shown in Fig. 4), and the detailed configuration is introduced in the supplementary (Section VII).

## Acknowledgements

The work was supported in part by the National Natural Science Foundation of China under Grants 61735006, 61631166003, 61675081, 61505060, 61320106016, and 61125501, in part by the China Postdoctoral Science Foundation under Grant 2018M640692, and in part by the Director Fund of WNLO.

## Author contributions

The experiment was designed and implemented by L. Chen, N. Yang, and X. Dong. The time-stretch microscope was co-developed by X. Zhou and Z. Lei, and the time-stretch OCT was co-developed by L. Zhang. C. Zhang and X. Zhang developed the concept and supervised measurements and analysis. All authors contribute the preparation of the manuscript.

## Competing financial interests

The authors declare no competing financial interests.

## Figure captions

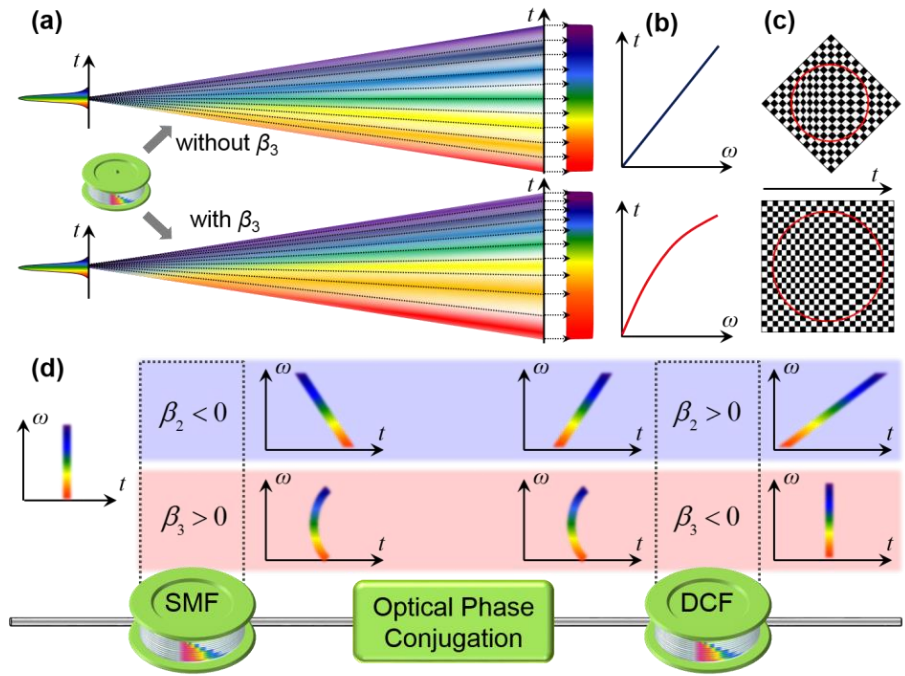

**Figure 1 | Conceptual idea of the pure temporal dispersion based on optical phase conjugation.** (**a**) Third-order dispersion introduced temporal asymmetry, compared with the quasi-linear $β_2$ during the dispersive time-stretch process, and resulted in the nonlinear frequency-to-time mapping (**b**). (**c**) Reconstructed two dimensional temporal aberration, only the horizontal temporal axis is distorted. (**d**) The effect of optical phase conjugation to $β_2$ and $β_3$ respectively, and $β_2$ with opposite values are accumulated while opposite $β_3$ is eliminated. Precisely length control between SMF and DCF can fully eliminate the third-order dispersion introduced quadratic temporal delay.

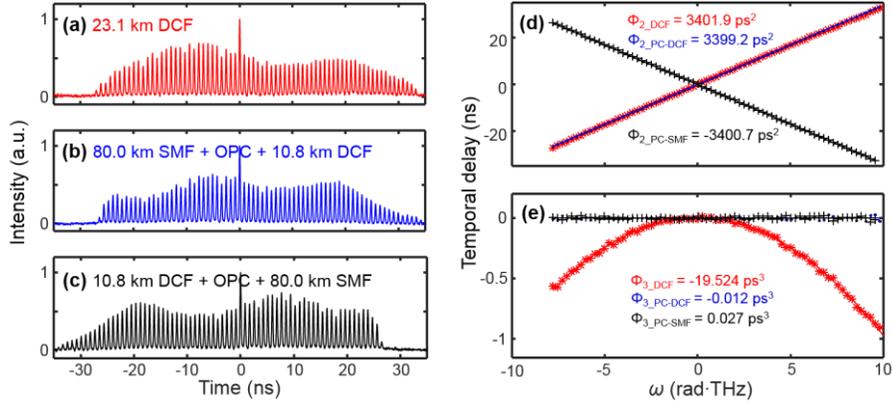

**Figure 2 | Characterization of the frequency-to-time mapping relation with 40-GHz equally spaced frequency comb lines.** **(a) – (c)** The dispersion elements under test include: uncompensated DCF (a); OPC with DCF afterwards to achieve normal dispersion (b); and OPC with SMF afterwards to achieve anomalous dispersion (c). **(d)** The overall temporal delay calculated from the relative temporal positions of each pulse peak, and normal dispersion (a & b) and anomalous dispersion (c) are observed. **(e)** The residual temporal delay by removing the second-order group delay dispersion. The uncompensated DCF exhibits obvious quadratic feature, while the OPC based pure temporal dispersions are almost zero.

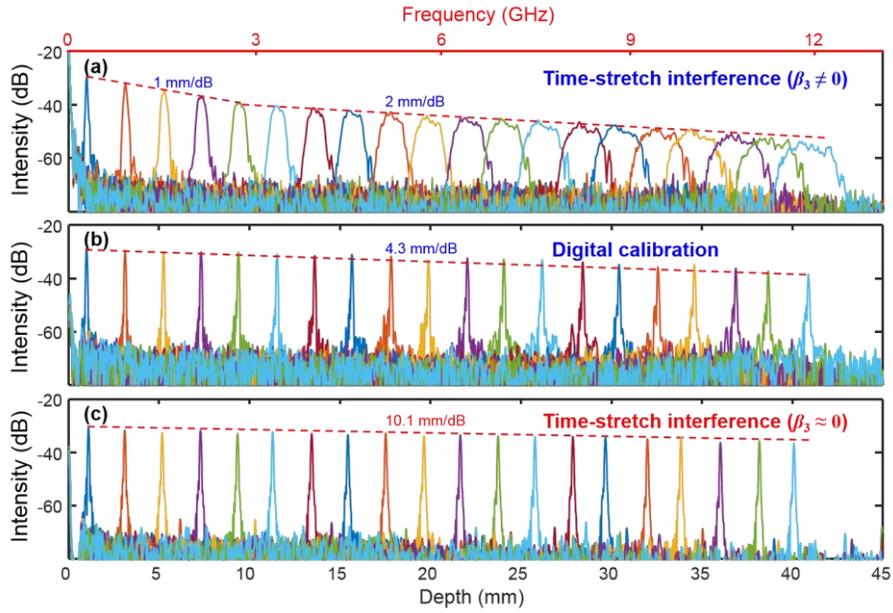

**Figure 3 | Roll-off curves of the scanned interference frequencies for different dispersive time-stretch swept-sources.** (**a**) With uncompensated third-order dispersion ($\beta_3 \neq 0$), and its roll-off slope is between 1~2 mm/dB. (**b**) The nonlinear swept-source in (a) is calibrated by the *k*-space resampling method (details see the supplementary information, part III). The roll-off curve is much flatter, and the spectral signal-to-noise ratio was enhanced. (**c**) With the OPC compensated third-order dispersion ($\beta_3 = 0$), and its roll-off slope is almost flat (~10.1 mm/dB). Obvious sensitivity enhancement is observed in the deeper depth range.

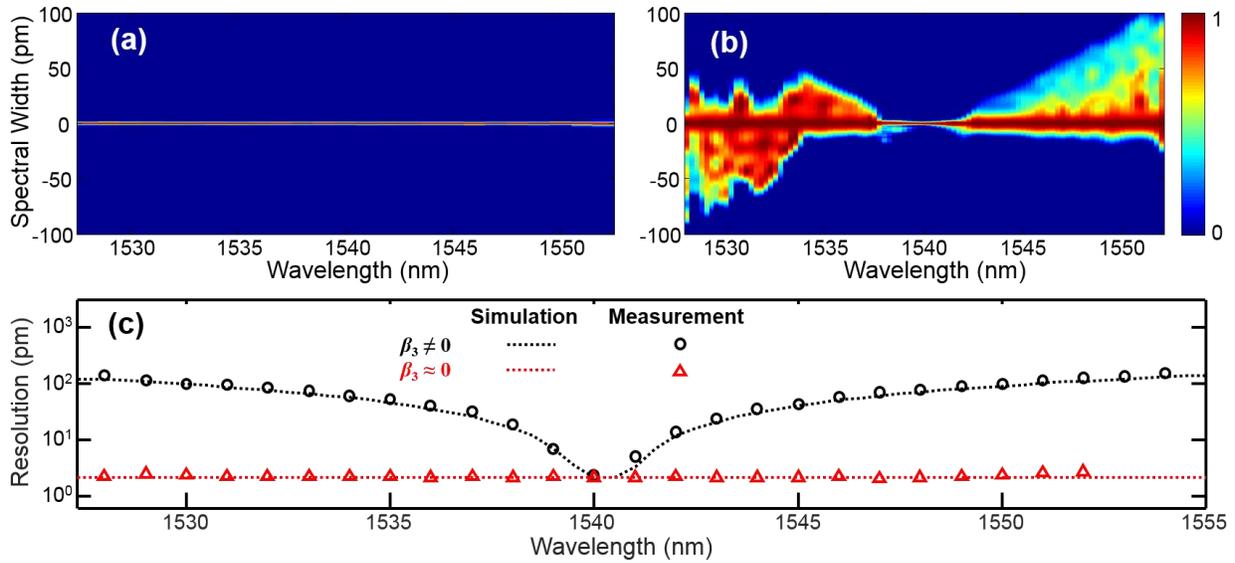

**Figure 4 | Resolution improvement of the PASTA system with pure temporal dispersion.** By tuning the signal under test from 1528 nm to 1552 nm, the measured spectral width in correspondence to the resolution is shown in **(a)** with pure temporal dispersion, and **(b)** with third-order dispersion. **(c)** The simulated (dotted line) and measured (black circle and red triangle) resolution degradations are figured out. It is observed that, for the 2 nm range around the central wavelength 1540 nm, these two configurations achieve similar resolution (~2 pm), while at the edge range (around 1530 nm or 1550 nm), the third-order dispersion degenerates the resolution from 2 pm to 0.1 nm.

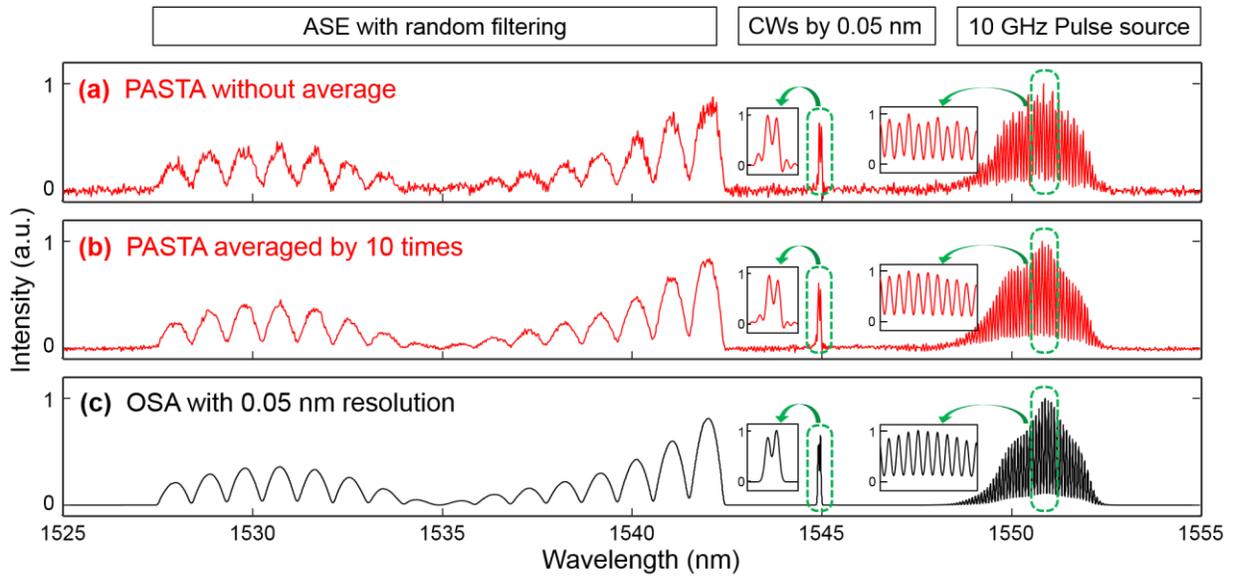

**Figure 5 | Wideband spectral-temporal analysis performance with pure temporal dispersion compared with conventional OSA.** Different spectral components are tested here, including ASE noise programmed by waveshaper, two CW components, and non-synchronized high repetition rate pulse source. The acquisition bandwidth of the PASTA system is 13 GHz, which is capable of achieving 50-pm spectral resolution. **(a)** Single-shot spectra captured by the PASTA with 20-MHz frame rate; **(b)** PASTA spectra averaged by 10 times to 2-MHz frame rate, and better spectral stability is achieved; **(c)** Spectra captured by the slow scanning (5 Hz) OSA with 0.05-nm resolution.

# Supplementary information

# Pure temporal dispersion for aberration free ultrafast time-stretch applications

Liao Chen, Xin Dong, Ningning Yang, Lei Zhang, Xi Zhou, Zihui Lei, Chi Zhang, [*] and Xinliang Zhang [**]

*Wuhan National Laboratory for Optoelectronics, Huazhong University of Science and Technology, Wuhan 430074, China*

*[*]chizheung@hust.edu.cn; [**] xlzhang@mail.hust.edu.cn*

**Section I ― Measurement of group delay dispersion**

Group delay dispersion (GDD) is the derivative of the group delay with respect to the angular frequency, and in a given optical fiber, the GDD per unit length can be expressed as $β_2$ in units of $s^2/m$. Plenty of methods have been developed to precisely measure this GDD of optical fiber, such as the pulse delay methods, the spectral interferometry for very small pieces of fiber or waveguide, the vector network analyzer (VNA) based phase-shift method, or some four-wave mixing (FWM) based method [S1-S4]. In this work, since large temporal dispersion (> 10 km dispersive fiber) was involved, which can be directly resolved in the temporal axis through the conventional detector and oscilloscope, thus a straightforward pulse delay (or time-of-flight) method has been introduced with its easy configuration shown in Fig. S1. Here, the broadband source centered at 1570 nm was provided by a mode-locked fiber laser with 20-MHz repetition rate and 25-nm bandwidth, and the equally spaced frequency components were filtered by a programmable optical filter (Finisar WaveShaper 100S) with 40-GHz separation. Passing through the dispersive fiber under test, the separated frequency components were stretched along the time axis, and their temporal positions can be directly obtained from a single-pixel detector and an oscilloscope with 13-GHz analog bandwidth, which determined ~200-ps$^2$ minimum total GDD according to 40-GHz frequency separation, and ~2.5-ps$^2$ GDD accuracy according to 25-nm spectral width. This pulse-delay manifested the GDD of the dispersive fiber, whose dispersive curve can be retrieved from the second-order polynomial fitting of the temporal positions. In this work, three types of dispersive configurations (with similar absolute dispersion of ~2.67 ns/nm) were measured with this pulse delay method: (a) 23.1-km DCF; (b) optical phase conjugation (OPC) with 80-km SMF in the front and 10.8-km DCF afterwards; (c) OPC with 10.8-km DCF in the front and 80-km SMF afterwards. The results were shown in Fig. 2 of the manuscript.

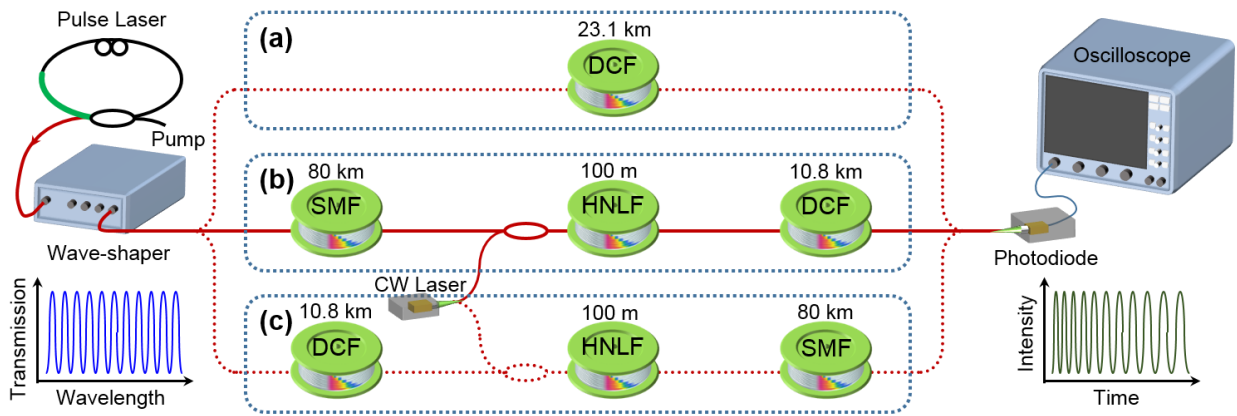

**Figure S1 | Schematic diagram of the group delay dispersion measurement.** (**a**) The normal dispersion is provided by 23.1-km DCF, with third-order dispersion. (**b**) The normal dispersion without third-order dispersion is provided by the OPC with 80-km SMF and 10.8-km DCF. (**c**) The anomalous dispersion without third-order dispersion is provided by the identical fibers with reversed position.

**Section II ─ Temporal deviation induced aberrations for time-stretch microscopy**

According to the pulse delay dispersion measurement, the third-order dispersion will induce the nonlinear frequency-to-time mapping, which results in the aberrations for the time-stretch spectroscopy and microscopy [S5,S6]. Among them, the aberrations of the microscope are easier to quantify through some recognized standard test targets, e.g. the 1951 USAF resolution target [S7]. Here, the spatial disperser of the time-stretch microscopy (space-to-frequency mapping) is configured identical as in Ref. [S8], and it achieved the spectral resolution of 0.18 nm and the corresponding spatial resolution of 2.3 μm. The deviation of the quasi-linear space-to-frequency mapping process is also measured by the aforementioned pulse delay methods, and it is observed from Fig. S2(a), up to 15-μm spatial deviation is observed from the central to the edge of the spectral band. This deviation is only observable in the horizontal direction, while the vertical direction acquired by the translation stages is almost linear, as discussed in Fig. 1(c) of the manuscript. To enhance the contrast of the minor spatial deviation, the images of the time-stretch microscopy with/without the third-order dispersion in different colors were overlapped, and their misalignment manifests the imaging aberrations as shown in Figs. S2(b)-(d). This aberration is similar when the test objects were rotated at different angles.

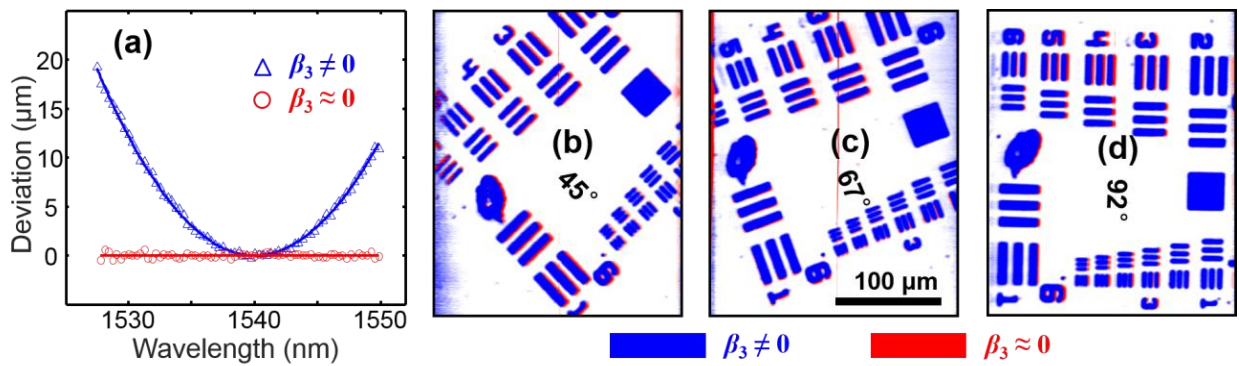

**Figure S2 | Aberration observation of the time-stretch microscopy with/without the third-order dispersion. (a)** The deviation of the quasi-linear space-to-frequency mapping

process, and it is mainly from the third-order dispersion ($\beta_3 \neq 0$). **(b)-(d)** The resolution test target (USAF 1951, Group 6 and 7) was introduced as the imaging object of the time-stretch microscopy, and rotated at different angle (45º, 67º, and 92º). Here, the blue and red colors represent the uncompensated and compensated third-order dispersion ($\beta_3$), respectively.

**Section III — Digital calibration for the interference frequency**

Due to the existence of the higher-order dispersion in the dispersive media, the dispersive time-stretch swept-source cannot achieve the precisely linear frequency-to-time mapping relation, which degraded the interference frequency with broadening bandwidth [S9]. Similar situation is also observed in the Fourier domain mode locking (FDML) based swept-source optical coherence tomography (SS-OCT), and results in degraded axial resolution [S10]. Therefore, a digital calibration process is usually required to retrieve the original frequency-to-time curve (black line) to the linear curve (red line), as the 1.5% deviation in Fig. S3(b). Here, the digital calibration process, also known as the *k*-space resampling, is performed by an interferometer with fixed optical path difference, and a chirped inference pattern is first obtained as the black line in Fig. S3(a) [S11]. The phase of the chirped fringe can be retrieved by the Hilbert transform, and it can further retrieve the nonlinear frequency-to-time curve. Then the next step is to generate a non-equally-spaced time axis based on the nonlinear curve, and resample the original inference pattern to obtain the linear frequency-to-time mapping as the red line in Fig. S3(a). Finally, the retrieved inference pattern is processed by the fast Fourier transformation (FFT) to obtain the undistorted axial information. Leveraging this digital calibration process, much flattened roll-off curve is achieved compared with original curve, as shown in Figs. 3(a) and (b) of the manuscript. It is noted that, the *k*-space resampling method has some hypotheses, such as the highly stable pulse-to-pulse repeatability and the ideal analog-to-digital conversion (ADC) process. However, in the practical application, the timing jitter of the pulse source and the limited quantization levels of the ADC process will degrade this calibration performance. Compare the roll-off performance between the digital calibration and the OPC compensation, the digital calibration results in lower sensitivity in the deeper depth range (Figs. 3(b) and (c) of the manuscript).

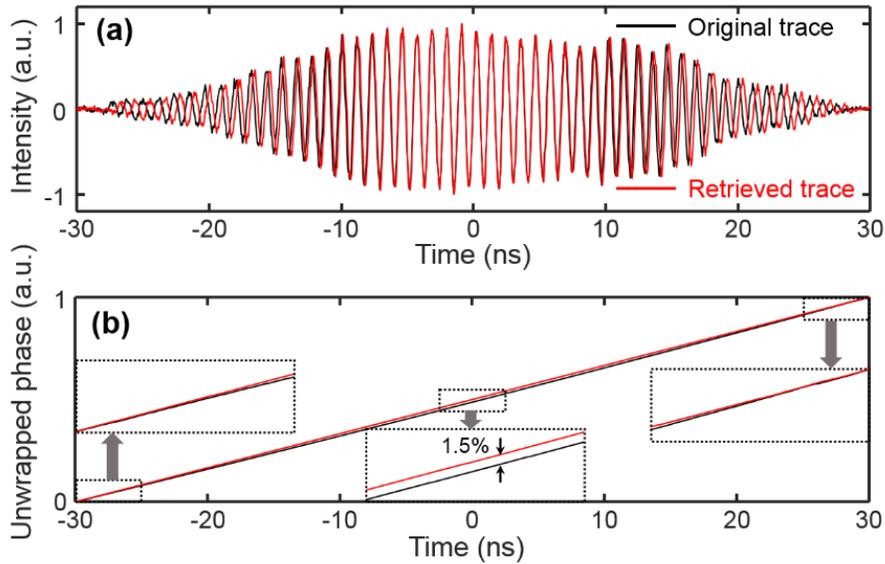

**Figure S3 | Digital calibration for the time-stretch based swept-source.** (a) The comparison of the original interference pattern (black solid line) and the retrieved interference pattern (red solid line), and the original trace shows increasing chirped frequency; (b) The comparison of the corresponding unwrapped phase curves.

**Section IV ─ Imaging depth enhancement for the OCT imaging system**

The roll-off curve manifests the imaging characteristics of the SS-OCT system, including the resolution, the sensitivity, and the imaging depth [S12]. To demonstrate the SS-OCT imaging quality improvement introduced by the OPC based optical calibration, the cornea and iris of a fish eye is introduced as the sample, and its depth is adjusted by moving the position of the reference arm. Figures S4(a) and (b) are the direct Fourier transformation of the interference pattern, thus their resolution is degraded as the increasing the depth, as shown in Fig. 3(a) of the manuscript. With the help of the digital calibration, the imaging performance is greatly improved as shown in Figs. S4(c) and (d), even though the fuzzy boundaries are observed with deeper depth of 38 mm in Fig. S4(d). In comparison, Figs. S4(e) and (f) are the images of the OPC based optical calibration, and there is no obvious changes as the sample moves deeper. To sum up, this SS-OCT system achieves 70-μm axial resolution with 15-nm spectral width, which matched well with the theory. The shallow sample is easier to achieve better performance, not only due to the less scattering, but also due to the better source coherence. Leverage the OPC based pure temporal dispersion, better imaging quality is achieved as the sample goes deeper, and no extra digital signal processing is required.

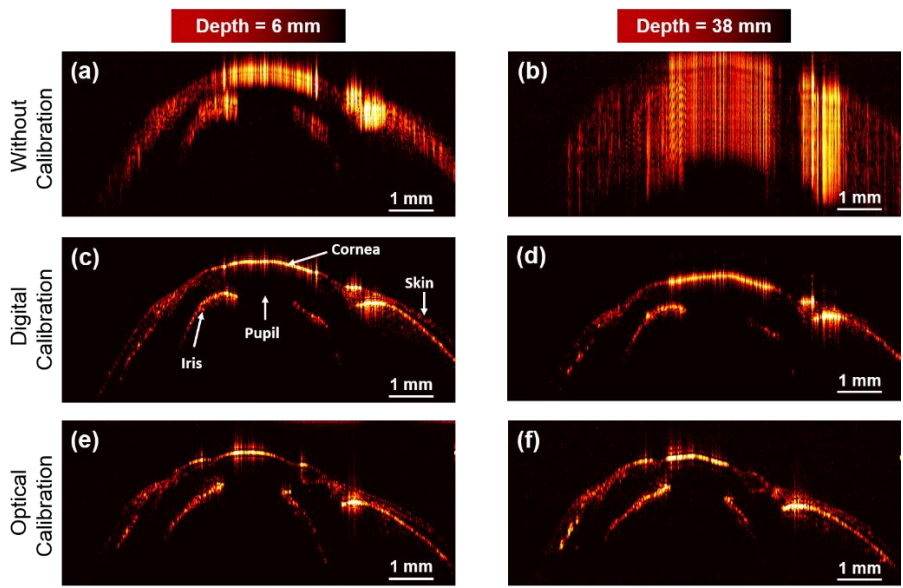

**Figure S4 | Swept-source optical coherence tomography images of the cornea and iris of the fish eye at different depths and calibration schemes.** Depth of 6 mm (a), (c), (e) and 38 mm (b), (d), (f) are performed by tuning the optical path length of the reference arm. (a) & (b) Direct

Fourier transform without calibration; (c) & (d) digital calibration based on the *k*-space resampling method; (e) & (f) optical calibration based on the OPC scheme.

**Section V ― Configuration of the compensated PASTA system**

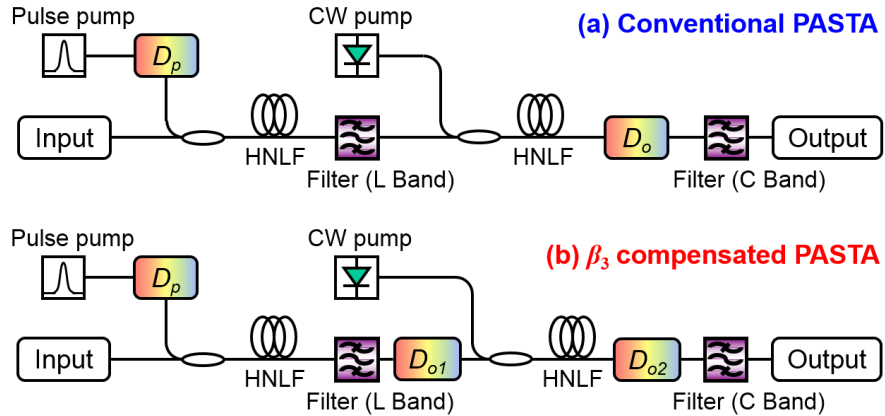

**Figure S5 | Configuration of the PASTA system.** (a) In the conventional PASTA system, the second-stage CW-pumped FWM is acted an optical phase conjugator, thus both the output dispersion ($D_o$) and the pump dispersion ($D_p$) can use identical type of DCF, which has large dispersion-to-loss ratio; (b) to compensate the higher-order dispersion of the output dispersion, it has been separated as two dispersive fiber $D_{o1}$ (SMF) and $D_{o2}$ (DCF), before and after the second-stage FWM. The other part of the configuration keeps identical.

Besides the direct time-stretch based absorption spectroscopy, arbitrary waveform can be also resolved through the ultrafast dispersive time-stretch, with the help of a converging time-lens, and configured as the parametric spectro-temporal analyzer (PASTA) [S13]. In the conventional PASTA configuration, there are two stages FWM involved, and the first one is acted as a converging time-lens, whose focal GDD is determined by that of the pump dispersion ($D_p$) [S14]. Since there are large dispersion involved (e.g. 2.96 ns/nm), to minimize the insertion loss of the system, the DCF with larger dispersion-to-loss ratio is preferable here than the SMF. Therefore, another stage FWM acted as an optical phase conjugator is applied, and the output dispersion can be also performed by the DCF, as shown in Fig. S5(a). However, the conventional PASTA suffers from limited observation bandwidth (e.g. 5-10 nm), which is mainly due to the higher-order dispersion of the output dispersive fiber. Leveraging the pure temporal dispersion introduced in this paper, the OPC process makes the SMF and DCF accumulate their $\beta_2$ and subtract their $\beta_3$. In the improved PASTA system, this pure temporal dispersion scheme replaces the original output dispersion, as shown in Fig. S5(b), and the original 12.89-km DCF based output dispersion ($D_o$) is separated as 44.44-km SMF ($D_{o1}$) before the OPC and 6-km DCF ($D_{o2}$) after the OPC. Most conventional SMF and DCF are applied here, though both of them have the third-order dispersion, with the dispersion coefficients as: $\beta_{S2} = -22.7$ ps$^2$/km, $\beta_{S3} = 0.113$ ps$^3$/km, $\beta_{D2} = 146.6$ ps$^2$/km, and $\beta_{D3} = -0.84$ ps$^3$/km, respectively.

**Section VI — Effects of the third-order dispersion for pump and output dispersion**

According to the aforementioned discussion, there are two spools of large temporal dispersion involved in the PASTA system, the pump dispersion ($D_p$, with the dispersion coefficient of $\beta_{2p}$ and $\beta_{3p}$) and output dispersion ($D_o$, with the dispersion coefficient of $\beta_{2o}$ and $\beta_{3o}$), and the pump frequency $\omega_p$ is regarded as the reference frequency $\omega_0$. The signal and idler frequency are $\omega_s$ and $\omega_i$ respectively, and to simplify the discussion, suppose $\omega_p = \omega_0 = 0$, thus $\omega_i = -\omega_s$. After the first parametric mixing with the swept pump, the frequency domain expression of the idler field can be approximated as:

$$\overline{E}_i(\omega) \approx \exp\left[-\ln 2 \frac{(\omega - \omega_i)^2}{\Delta \omega_p^2}\right] \exp\left[-i \frac{\beta_{2p} L_p}{4}(\omega - \omega_i)^2 - i \frac{\beta_{3p} L_p}{4}(\omega - \omega_i)^3\right] \quad (S1),$$

where $\Delta\omega_p$ is the bandwidth of the swept pump. The frequency domain transfer function of the output dispersion can be expressed as:

$$G_o(\omega) = \exp\left[-i\frac{\beta_{2o}L_o}{2}\omega^2 - i\frac{\beta_{3o}L_o}{2}\omega^3\right] \quad (S2).$$

Therefore, we can have the output field expression in the frequency domain:

$$\begin{aligned}
\overline{E}_o(\omega) &= \overline{E}_i(\omega)G_o(\omega) \\
&\approx \exp\left[-\ln 2 \frac{(\omega-\omega_i)^2}{\Delta\omega_p^2}\right]\exp\left[-i\frac{\beta_{2p}L_p}{4}(\omega-\omega_i)^2 - i\frac{\beta_{3p}L_p}{4}(\omega-\omega_i)^3 - i\frac{\beta_{2o}L_o}{2}\omega^2 - i\frac{\beta_{3o}L_o}{2}\omega^3\right] \\
&= \exp\left[-\ln 2 \frac{(\omega-\omega_i)^2}{\Delta\omega_p^2}\right]\exp\left[-i\Phi_3'(\omega-\omega_i)^3 - i\Phi_2'(\omega-\omega_i)^2 - i\Phi_1'(\omega-\omega_i) - i\Phi_0'\right]
\end{aligned}$$

(S3).

Here, four newly defined coefficients $\Phi'_3$, $\Phi'_2$, $\Phi'_1$ and $\Phi'_0$ can be expressed as:

$$\begin{cases}
\Phi_3' = \dfrac{\beta_{3p}L_p + 2\beta_{3o}L_o}{4} \\[4pt]
\Phi_2' = \dfrac{\beta_{2p}L_p + 2\beta_{2o}L_o + 6\beta_{3o}L_o\omega_i}{4} = \dfrac{\beta_{2p}L_p + 2\beta_{2o}L_o - 6\beta_{3o}L_o\omega_s}{4} \\[4pt]
\Phi_1' = \dfrac{3\beta_{3o}L_o\omega_i^2 + 2\beta_{2o}L_o\omega_i}{2} = \dfrac{3\beta_{3o}L_o\omega_s^2 - 2\beta_{2o}L_o\omega_s}{2} \\[4pt]
\Phi_0' = \dfrac{\beta_{2o}L_o\omega_i^2 + \beta_{3o}L_o\omega_i^3}{2} = \dfrac{\beta_{2o}L_o\omega_s^2 - \beta_{3o}L_o\omega_s^3}{2}
\end{cases} \quad (S4),$$

and they can fully describe the degradation of the output field. To simplify the discussion, these four parameters are analyzed separately. Firstly, for the third-order term $\Phi'_3$, it is the accumulated third-order dispersion from the pump dispersion to the output dispersion, and results in an oscillating trailing edge for the output pulse (also affected by $\Delta\omega_p$), see the central inset of the Fig. S6(a) [S15]. Secondly, for the second-order term $\Phi'_2$, it is the accumulated third-order dispersion, and results in the broadening for the output pulse. It is noted that, $\Phi'_2$ is an input frequency ($\omega_s$) dependent coefficient, as long as the third-order dispersion of the output dispersion ($\beta_{3o}$) is uncompensated, also see the insets of the Fig. S6(a). Thirdly, for the first-order term Φ'$_1$, it is the temporal shift term, and determines the frequency-to-time mapping relation of the PASTA system. Ideally, the linear frequency-to-time mapping is performed by the term $β_{3o}L_oω_s$, while the third-order dispersion ($β_{3o}$) introduces nonlinearity here, as shown in Fig. S6(a), where the black pulse positions are not precisely aligned with the grids. Lastly, the zero-order term Φ'$_0$ is a constant phase shift term, and it has no effects for the output pulse intensity.

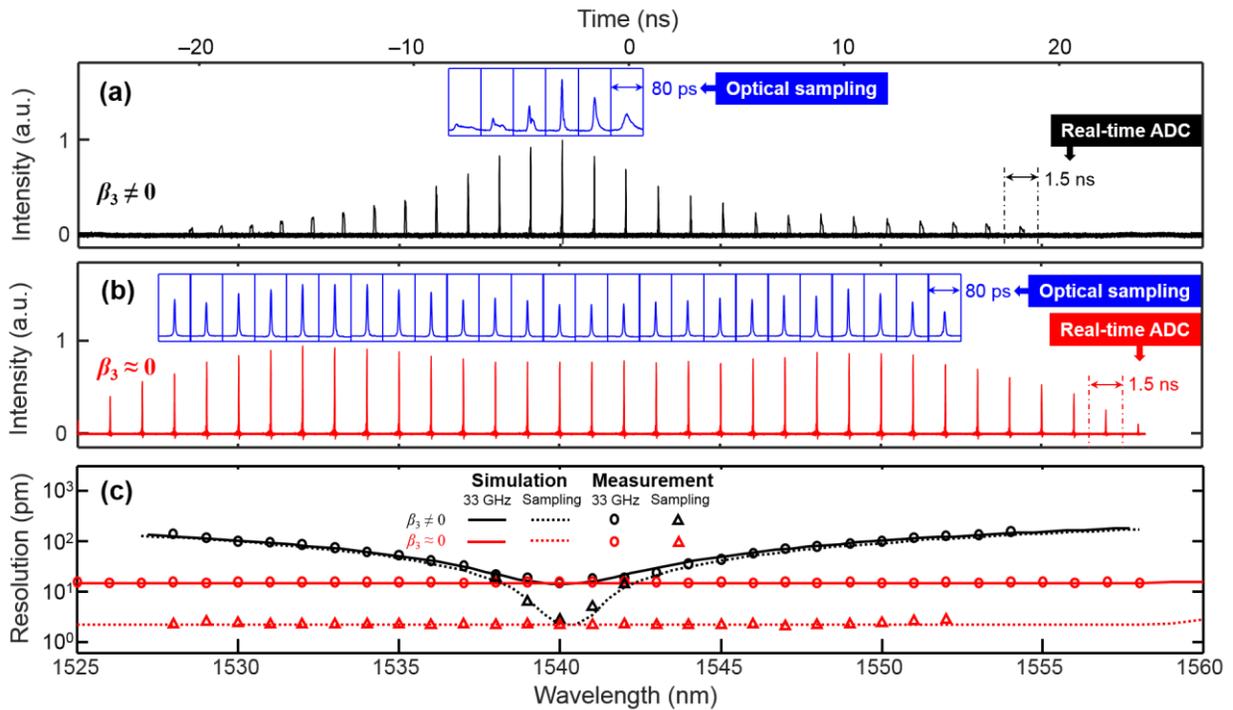

**Figure S6 | Observation bandwidth and resolution of the PASTA system.** (**a**) Due to the third-order dispersion of the output dispersion, the resolution across the observation bandwidth degrades very fast, and the wavelength-to-time mapping relation is a little grifting. Black line: 33-GHz real-time acquisition bandwidth; blue line: optical sampling scheme. (**b**) When the third-order dispersion is compensated, the spectral width across the 30-nm observation range keeps identical, no matter the 33-GHz real-time acquisition bandwidth (red line) or the optical sampling (blue line). (**c**) Resolution performance with/without the third-order dispersion

compensation, which is another manifestation for the Fig. 4 of the manuscript. In general, larger acquisition bandwidth benefits sharper resolution, and the third-order dispersion determines the observation bandwidth, which aligned with the simulation results.

Leveraging the OPC scheme introduced in this paper, the third-order dispersion of the output dispersion is fully compensated ($\beta_{3o} = 0$). As a result, the $\Phi'_3$ is roughly reduced by half, and the oscillating trailing edge would be smaller; the $\Phi'_2$ can be fully compensated to zero across the whole observation bandwidth, and maintains the output pulsewidth and resolution constant as shown in Fig. S6(b); the temporal shift term $\Phi'_1$ becomes fully linear, and enhance the spectral accuracy of the system. Therefore, this higher-order dispersion compensation scheme has greatly improved the PASTA performance, including the observation bandwidth, resolution, and accuracy.

**Section VII ─ Dual optical frequency combs based asynchronous optical sampling (ASOPS)**

To improve the resolution of the PASTA system, large acquisition bandwidth is highly desired; however, the state-of-art high-end real-time photodiode and ADC can only achieve around 100-GHz bandwidth, which is not enough to resolve those sub-picosecond pulses [S16,S17]. The pump-probe technology based on optical sampling provides an economical solution to resolve those ultrashort pulses, though its frame-rate is sacrificed. Here, dual optical frequency combs were introduced with slightly different repetition rates $f_r$ and $f_r + \Delta f$, which resulted in pulse-to-pulse walk-off ($\Delta t = \Delta f/f_r^2$) between two combs in time domain, and the time axis was scaled up by a factor of $M = f_r /\Delta f$. As a result, the acquisition bandwidth requirement was thus reduced by the same factor. In general, the dual optical frequency combs were required to have stabilized carrier frequency as well as the repetition rate, especially for the interferometer, such as the dual-comb spectroscopy [S18]. Otherwise, the free running carrier frequencies would result in the frequency jitter. To simplify the requirement and avoid the locking of the carrier frequencies, directly extracting the intensity profile was carried out by a nonlinear optical sampling process, namely the nonlinear ASOPS process [S19]. In this paper, the ASOPS was performed based on the parametric FWM process, where the phase information was filtered out by the square law detection of the idler field. The repetition rates of the two frequency combs were 92.07 MHz and 92.074 MHz, respectively, and the magnification factor of $M = 23000$ was achieved here. In the time domain, different idler pulses carried the scanned envelope of the signal, presenting the sampling process. Moreover, owing to the low acquisition bandwidth requirement, improved sensitivity was also achieved.